\documentclass[showpacs,showkeys,preprintnumbers,amsmath,amssymb,superscriptaddress]{revtex4}

\usepackage{graphicx}
\usepackage{dcolumn}
\usepackage{bm}
\begin{document}

\pagestyle{myheadings}

\title{Osmosis at constant volume
 and  water uptake in tall trees   }

\author{Pa\v{s}ko \v{Z}upanovi\'{c}}
 \email{pasko@pmfst.hr}
\affiliation{ University of Split, Faculty of  Science\\  Teslina 12,  21000 Split,  Croatia }
\author{Milan Brumen}
\email{Milan.Brumen@uni-mb.si }
\author{Ale\v{s} Fajmut}
\email{ales.fajmut@uni-mb.si}
\affiliation{Faculty of Natural Sciences and Mathematics, Faculty of Medicine and Faculty of Health Sciences, 2000 Maribor, and Jo\v{z}ef Stefan Institute, 1000 Ljubljana, Slovenia}
\author{Domagoj Kui\'{c}}
 \email{dkuic@pmfst.hr}
\author{Davor Jureti\'{c}}
 \email{juretic@pmfst.hr}
\affiliation{ University of Split, Faculty of  Science\\  Teslina 12,  21000 Split,  Croatia }

\date{\today}




\begin{abstract}
 We consider a  thermodynamic state  of a solvent and solution separated with an  elastic semipermeable membrane in a box with a constant  volume  and the 
  relevance of this simple model for the water uptake in tall trees.
Under moderate concentrations of  a solute, the  solution and  solvent are under the positive and negative pressure, respectively. 
  In the case of the soft membrane  the pressure 
  difference between  the compartments with the solvent and solution is given  by van't Hoff equation. A state of the negative pressure is not stable and after some time 
   cavitations transform the solvent into the state of coexisting liquid and  bubbles of saturated vapor.
       The pressure difference between the solvent and solution decreases and the membrane relaxes restoring the  liquid phase in the compartment with solvent. In this way the solvent oscillates between the tensile  state and  the coexisting  state of 
      liquid and bubbles of saturated vapor. 
      
      The xylem and phloem, the main vascular systems in trees, are   coupled with ray cells.  Assuming that a sap in these  systems is kept under the constant volume the osmosis 
       between the xylem and  phloem with  ray cells sustains the negative pressure of the  xylem sap. 
       Due to the osmosis   elastic energy stored in walls of tracheary element  
        could repair cavitations.
      In this way both water transport in tall trees and cavitations repair in  tracheary elements are related to the osmosis under  constant volume. 
       A possible explanation of two  long standing  problems in tree physiology,  
     water uptake in tall trees and cavitations repair, are offered within this model.
\end{abstract}

\maketitle

\section{Introduction}
Water, and generally liquids, is  hardly  compressed  if the corresponding state is not too close to critical point. The usual values for bulk modulus, inverse compressibility, of liquids are of the order of  $GPa \approx  10^4 p_a$, where $p_a$ is the standard atmospheric pressure.   On the other hand if the liquid is exposed to the tension one would expect, following standard thermodynamic books \cite{zemansky,kubo,guggenheim}, that the tension  equal to the difference between the  ambient pressure and  the pressure  of saturated vapor at the given temperature would be enough to detach the piston from the column of water. However, it is hard to believe that there is such discontinuity in the compressibility. Indeed, experiments show that the column of  water
 in the cylinder closed with a piston  behaves like a solid   adhesively bound to the inner surface of the cylinder and piston. One needs to apply   0.13 GPa in order to break the  water column under the standard  conditions. It is usually said  that the tensile  liquid is the liquid   under  negative pressure.
 
 The state of the liquid under moderate  negative pressure is metastable. Vapor  bubbles in  liquids   generated by fluctuations, either on the contact surface between fluid and container or within the liquid
  \cite{atilla0}, are the seed of  the  phase in which  vapor bubbles and liquid coexist. However if a  bubble 
  does not  achieve a critical  radius the increase of the pressure within it, due to the surface tension, will convert it  back into the liquid.  
 The radius of the vapor bubble is a statistical quantity  and the tensile liquid after some time  turns into the stable phase of the coexisting  liquid and vapor. The higher the tension applied to liquid the shorter will be the lifetime of the   metastable state \cite{atilla}.  
 Once the  tensile liquid turns into stable phase of coexisting  liquid and vapor 
  the pressure necessary to  expand (or compress) system  is equal to the difference between the ambient pressure and the pressure of saturated vapor. 
 
  Liquids under  negative pressure have been  considered for a long period of time. The first experimental observation was reported by Huygens in 1661-62 \cite{kell}. 
   Liquids under negative pressure have come into focus after publication of  Dixon and Joly theory (1984)  about  water uptake in tall trees known as the cohesion tension (CT) theory  \cite{dixon}.
  Recently  the interest in liquids under negative pressure  has been revived
    \cite{atilla,atilla1,tdh,dpg}. The state of the  negative pressure has been found in  liquid systems \cite{atilla,atilla1} and  liquid-liquid systems \cite{atilla21,atilla3}. Although it is well known fact that the osmosis under constant volume leads to the state of the  negative pressure
  \cite{hirsch1,freeman,hirsch2,Scholander} there is no, at least to our knowledge, quantitative analysis of this effect.
  
  In the first part of this paper  we consider osmosis in a box of constant volume divided by the  semipermeable membrane into two compartments. 
  Relevant thermodynamic quantities, 
   free energy change of the solvent due to the transition of solvent molecules from compartment with the solvent to compartment with the solution, elastic energy of the membrane and configuration entropy of the solute are kept at equal footing. The minimum of free energy determines pressures in compartments. 
   A moderate concentration of solute induces the negative pressure in the compartment with solvent.

  In the second part of the paper we revisit the problem of water uptake in tall trees and apply results derived in the first part to the main  vascular systems in trees.  
 These are the  xylem and phloem \cite{PlantPhysiology}.
   Water with some minerals and enzymes (xylem sap) flow  upward through the  xylem, system of nonliving cells. A solution  rich with products of  photosynthesis (phloem sap) flows through 
the phloem, system of living cells \cite{2amerikancabook}. Depending on the season stored photosynthesis products or their derivatives flow downward or upward. 
The xylem and phloem are connected with  ray cells that serve mainly as storage of photosynthetic products   and/or their derivates.
We call the  xylem and  phloem  with ray cells the extended vascular system.

   It is widely accepted that the CT theory explains long distance water transport  although some authors had expressed  skepticism regarding its validity \cite{Zimmerman,Canny}. 
    According to the CT theory the driving force for water movement in the system is generated by the surface tension. 
    Evaporation generates a curvature in the water menisci within the pores of the  mesophyll cell wall   lined by cellulose microfibrils.
 The radius of curvature of these menisci is sufficiently small as to be able to support water columns as long as one hundred meters that is an approximate height of the tallest existing trees
  \cite{Cruiziat}. Close to the ground the   xylem sap is approximately  under standard atmospheric pressure ($0.1 MPa$).
  In order to lift water  up to the top of one hundred meter high tree the sap pressure close to the top should be less than   $-0.9 MPa$.
  Water under the  high tension  is in a metastable state \cite{atilla1} and it is susceptible to  cavitate into the  coexisting phase of  water and   saturated vapor.  
   Cavitations spread over the whole noninterrupted liquid  volume. If this were a case in trees  the ascent of sap in xylem  would be terminated leading to the withering of a tree. 
   However, the xylem is compartmentalized into tracheary  elements  and cavitations are usually isolated within one tracheary  element.

\begin{figure}[ht]
\centering
\resizebox{14cm}{16cm}
{\includegraphics{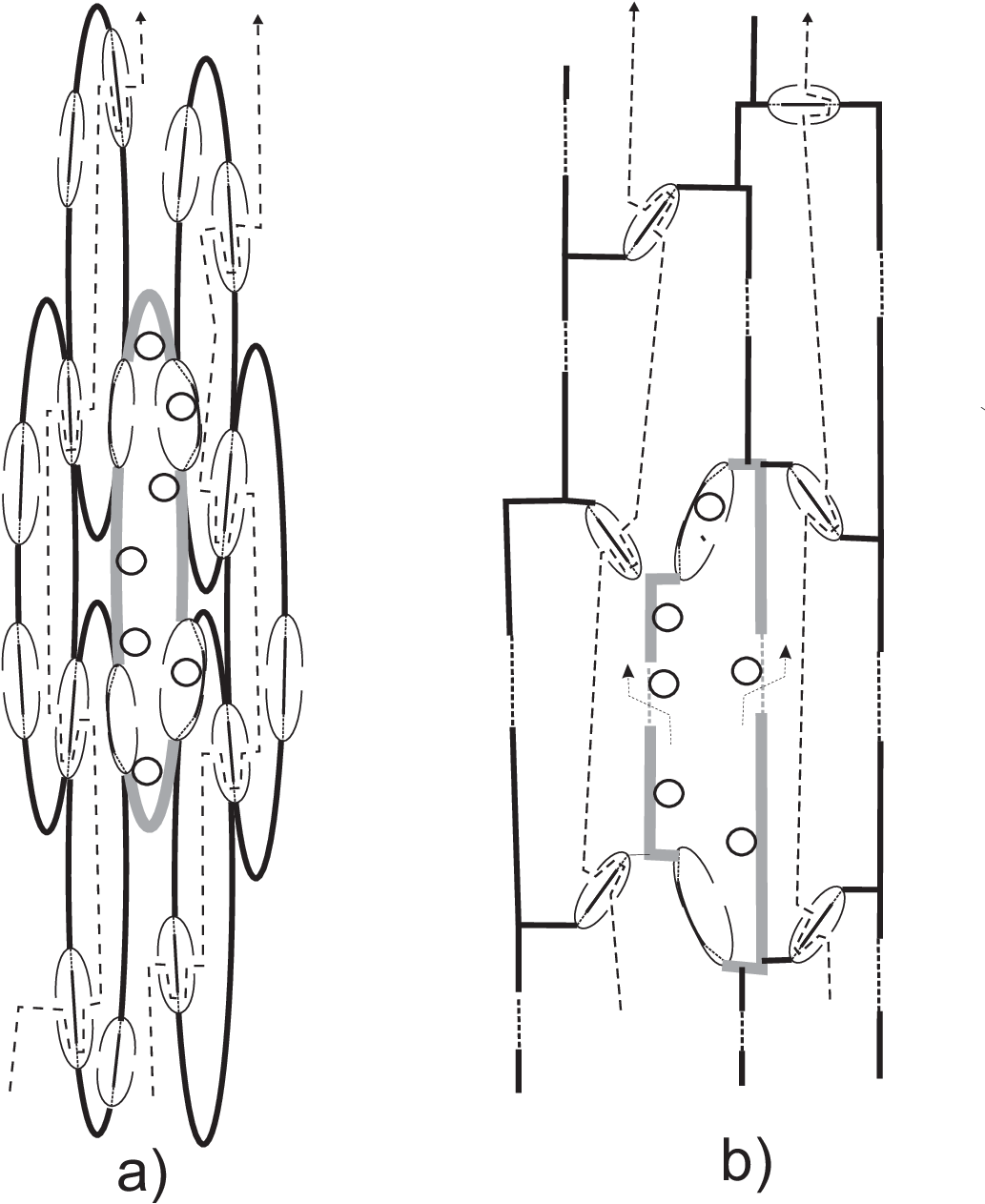}}
	\caption{\label{Figure1}  A cavitated tracheid (a) and  vessel (b)  surrounded by ones under the negative pressure. Gray walls belong to  cavitated tracheary elements.  Bubbles of saturated vapor are depicted as circles sticked to inner walls of the tracheary element. Dashed lines show  directions of water fluxes. The xylem parenchyma is not depicted in this figure.}
\end{figure}

    There are two classes of tracheary elements involved in the xylem water transport, tracheids and vessels. Tracheids are present in both  angiosperms and  gymnosperms, as well as in ferns and other group of vascular plants.     
    Vessel elements are found in angiosperms, a small group of gymnosperms called the Gnetales, and some ferns. 
     Tracheids are elongated hollow nonliving cells with lignified walls. There are several types of tracheids \cite{PlantPhysiology}.
      In angiosperms nonliving cells called vessel elements stack forming a pipe like structure of finite length called a vessel.  In this paper we consider tracheids in conifers and vessels in angiosperms. Tracheids in conifers are interconnected through numerous numbers of bordered pits [17]. Bordered pits act like a valve (see Fig.\ref{Figure1}a).
    End walls  of vessels (those that are inclined to the  horizontal plane) overlap.  In overlapping areas vessels are hydraulically connected by bordered pits.  In contrast to tracheids of conifers 
     side walls of vessels are interconnected with  pits, a membrane with small pores built in an aperture between neighboring vessels (see Figure \ref{Figure1}b).
      In this way tracheids and vessels make hydraulic networks in  a tree. 
    
    Cavitations set up  a pressure difference    on the pit membrane  torus  that closes the bordered pit  \cite{Tyree} disabling water transport through the pit channel. 
    In this way the cavitated  tracheid   is isolated from the neighboring tracheids with the sap under the tension
      (see Fig.\ref{Figure1}a). 
      In  vessels closed bordered pits disable a vertical water transport through cavitated vessel.
     Although pits in side walls are not closed, pores in   pit membranes are so small (few nanometers according to reference \cite{Choat} ) and very high pressure difference between vessels is necessary to drag a meniscus that is  formed as interface between vapor in  cavitated  vessel   and tensile sap in neighboring ones. 
       Still there is a radial water flow through pits in side walls (see Fig. \ref{Figure1}b). Due to small pores in the pit membrane there is small hydraulic conductivity and small water flow from cavitated vessel to the low pressure part of  the xylem. The radial water flow     does not 
       spoil tension state of  the sap in  the neighboring vessels. 
    In these ways  cavitations are  localized within the  tracheary element. The cavitated tracheary element slightly decreases    xylem hydraulic conductivity    
        but due to  interconnections between  tracheary elements  water supply to the top of the tree is not interrupted (see Figs.\ref{Figure1}a and \ref{Figure1}b).

      If there were no process reverse to cavitations cavitated vessels  would  be disabled to take a part in water transport.  
     In the course of time the sap in all vessels within one cross section  would undergo cavitations  leading to the withered  trees. However this is not a case. 
     In essay \cite{essay}   that accompanies recently published book \cite{PlantPhysiology} J. K. Wheeler and N. M. Holbrook   have  divided  experimental results  that 
     are possible proof of    vessel  refilling into two categories (see  \cite{essay} and references therein. Experiments from the first group directly measure the sap pressure in the xylem.
      These methods  
       lead to the  destruction of the sample with a possible  creation of  artifacts. Another group of  experiments are studies in which plants are subjected to  treatments designed to alter physiological parameters that reduce the degree of refilling,   
         such as phloem transport (girdling), starch reserves (shading), or membrane activities (addition of sodium orthovanadate or fusicoccin). These results  are  then compared with those obtained from  control plants. They believe   that  latter studies are the strongest evidence of the tracheary  elements refilling under the tension.  
     According to  J. K. Wheeler and N. M. Holbrook \cite{essay} 
     these "perturbation" studies have led to the conclusion that the metabolic source of energy, most likely sugars  stored in the ray cells as starch or supplied by the phloem, 
     is required if 
      refilling under tension occurs. The phloem should be the source of water supply.
      
       Recently the high-resolution computed tomography has been used to visualize, in vivo, refilling of  vessels \cite{Brodersen}. This paper clearly shows that  
       vessel refilling under tension  does exist.

       There are many proposed  mechanisms for vessels  refilling  under  negative pressure  \cite{Clearwater,essay}. Some of these proposals assume existence of 
        solute molecules in the cavitated  tracheary  element. The osmosis is then invoked as the possible mechanism of the vessel refilling  \cite{2amerikanca2,Salleo,Tyree}. Other theories  invoke an additional pressure in the 
         xylem surrounding tissue that should be responsible for tracheary  element refilling \cite{Canny,Bucci}.
         Proposed mechanisms remain, to a large degree, unsubstantiated \cite{essay}. 
          We refer the reader interested in these and other  theories to the review paper written by M.J. Clearwater and G. Goldstein in reference   \cite{Clearwater} and  references therein.
           In spite of numerous attempts to find the mechanism of the xylem refilling under tension this is still an open question.
            However, there is a common denominator among these attempts. It is generally accepted that surrounding tissue, the  phloem with ray cells  should be taken into account in order to explain
             this effect.

    Cavitations are not only mechanism that breaks the tensile state of the  xylem sap. Tracheary  elements are connected with neighboring ray cells (xylem parenchyma) by pits. 
    The pressure difference on the pit membrane can suck a gas bubble  sticked on  the xylem parenchyma side of the pit membrane. Once the gas bubble finds itself in the tracheary  element it rapidly expands breaking the tensile state of sap. This phenomenon is  called air seeding (see reference \cite{Tyree19954}). Malfunction  of  tracheary  elements due to the gas presence 
     is called embolism. Cavitations and embolism are two sources of interruption of the water uptake in trees.   In this paper we consider cavitations and propose a possible 
      mechanism  of the tracheary  elements  refilling.

  The paper consists of eight sections. After this introductory section we consider, in the Section \ref{2}, the osmosis under constant volume. Elastic energy of the membrane is added to the standard description of the osmosis. Due to the osmosis the solvent is under the negative pressure and after a while cavitations occur. Cavitations relax the elastic
membrane. Elastic energy stored in the membrane sustains the solvent flux from the solution to the compartment with the solvent. Once the liquid phase is restored in the solvent compartments the osmotic process starts again. The gravitation effect is considered in the Section \ref{tube}. Modifications of containers that do not affect the osmosis under constant volume are subjects of the Section \ref{4}. The role of the elasticity of the membrane in the process of cavitations repair is elaborated in Section \ref{5}. In the Section \ref{xylem} the theory developed in previous sections is applied to the problem of water uptake in tall trees and the refilling of tracheary elements under the tension. In the Section \ref{SecondLaw} the proposed mechanism for water uptake and refilling of vessels is put into relation with the second law of thermodynamics.
Finally results are summarized in Conclusions.

\section{Osmosis under constant volume }
\label{2}
 
\begin{figure}[ht]
\centering
\resizebox{16cm}{12cm}
{\includegraphics{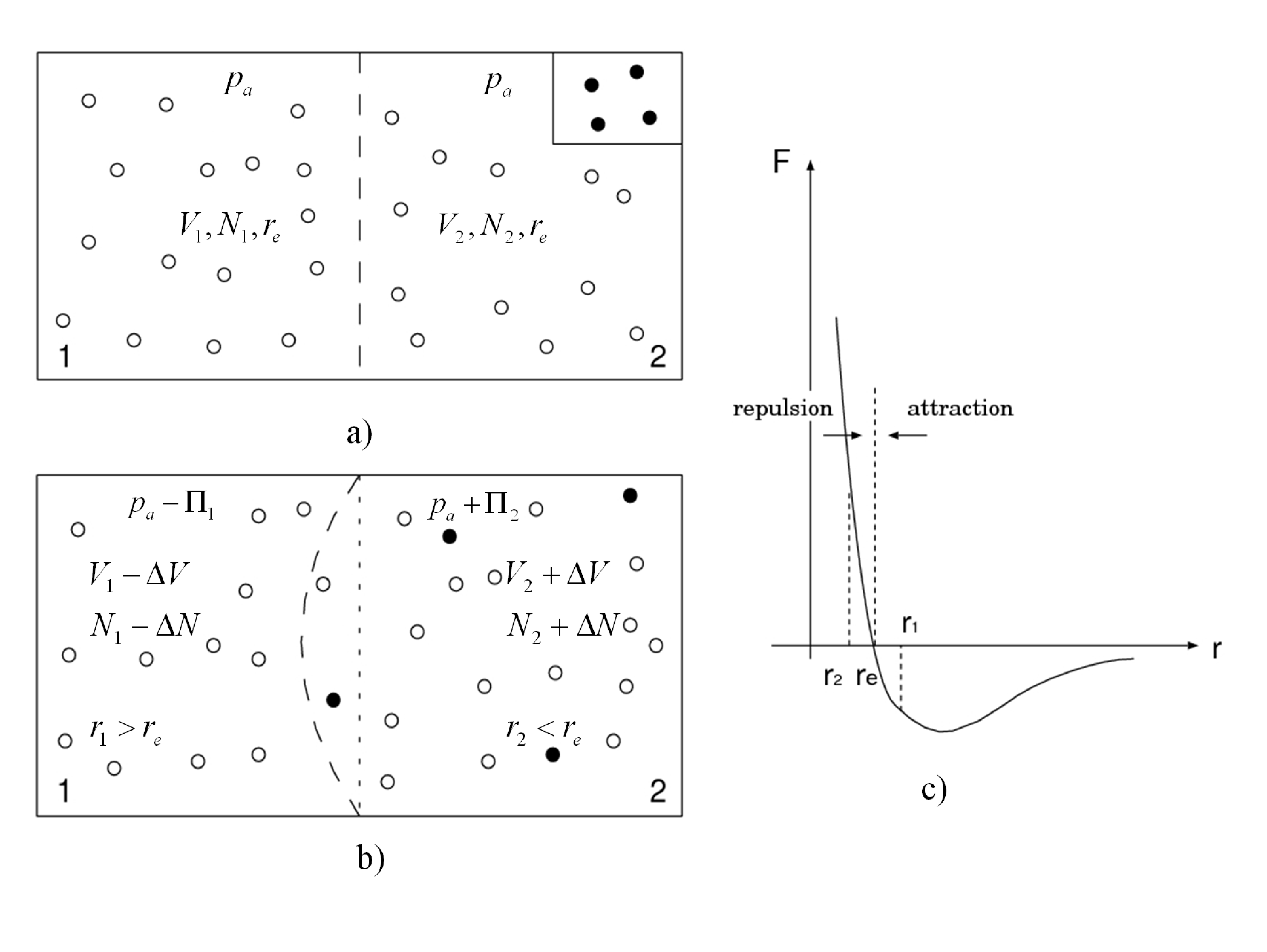}}
	\caption{\label{Figure2} 
	 The schematic representation of the formation of osmosis. a) The solute  (black circles) separated from the solvent  (white circles). b) The solute  released in the compartment. c) The qualitative graph of the intermolecular forces. }
\end{figure}

 In this section we consider the osmosis within the box  separated in two compartments by the  membrane impermeable for solute molecules.  After removal of  walls  
  between solute and solvent  (see Fig. \ref{Figure2}a)   two processes start,  diffusion of solute into the compartment with solvent and  the osmosis between the compartments. 
 A relaxation time of diffusion is much shorter than one of osmosis and 
 we can take that osmosis (see Fig. \ref{Figure2}b)  follows  diffusion. 
 
In order to describe osmosis quantitatively we define the  
mean  distance between molecules as  the  third root of 
  the inverse value of the concentration,
\begin{equation}
\label{md}
r=\sqrt[3]{\frac{V}{N}}. 
\end{equation}
 Further,   without loss of generality, we assume, that solvent is under  atmospheric pressure $p_a$ before the osmotic process has started (see  Fig. \ref{Figure2}a). We call the solvent  under these conditions free solvent.
 The mean distance between molecules in free solvent is $r_e$  (see Fig. \ref{Figure2}a and \ref{Figure2}c). We call $r_e$  equilibrium distance between solvent molecules. 
 
 We take, for the sake of convenience,   that solvent is water.  The solution is mixture of water
molecules and solute molecules. 
   Transition of   water molecules from  water into solution is  accompanied,  due the constant volume of the box,  by the 
  increase and decrease   of the  mean distances between  molecules, relatively to equilibrium distance   $r_e$, in water and solution  (Fig. \ref{Figure2}b), respectively.
  The pressure is a decreasing function of mean distance and it decreases  in compartment with water.
 At some critical value of the concentration of  the solute the  pressure in the  compartment with water falls to zero. Further increase of the  concentration leads to the  tensile state of water. It is usually described as the state of  negative pressure.  
 In accord with the    Le Chatellier-Brown principle the
     membrane deforms   increasing the volume of the  compartment with solution  (Fig. \ref{Figure2}b).  The result is  the membrane    under tension compressing solution  and expanding  water, respectively.
     In the terms  of the standard   intermolecular force  (Fig. \ref{Figure2}c) the intermolecular distance between molecules in left and right compartments are   $r_1>r_e$ and   $r_2<r_e$, leading to the attractive and repulsive force among  molecules
     in water and solution, respectively.

     The change of free energy of the system is a sum of the change of free energy of water in both compartments,  elastic energy stored in the membrane and increase of  entropy due to the presence of solute.
     The next three subsections are devoted to the calculation of these   contributions to  free energy.

     \subsection{Change of  water free energy }
     \label{a}

 Due to the small compressibility of water, at moderate tensions, change of water free energy  can be described by  Taylor expansion in terms of the change of mean molecular distance from its equilibrium value $r_e$,

 \begin{equation}
{\cal F}_i(r_i) = {\cal F}_i(r_e) + \left. \frac{\partial {\cal F}_i}{\partial r_i}\right|_{r_i = r_e}(r_i - r_e) + \frac{1}{2}\left. \frac{\partial^2 {\cal F}_i}{\partial r_i^2}\right|_{r_i = r_e}(r_i - r_e)^2.
\end{equation}
Exploiting the relationship between molecular concentration and the mean molecular distance   (\ref{md})
derivations of free energy in the foregoing equations   can be written in terms of  pressure and the  compressibility, 
\begin{equation}
\label{deltaF1}
\Delta {\cal F}_i(r_i) = 3V_{ei} p_i \left (\frac{r_i}{r_e} - 1\right) + \frac{1}{2}\left [\frac{9V_{ei}}{\kappa} + 6V_{ei} p_i \right ]\left (\frac{r_i}{r_e} - 1\right)^2.
\end{equation}
Here 
\begin{equation}
	p_i=-\frac{\partial {\cal F}_i}{ \partial V_i}
\end{equation}
is the pressure in the $i-th$ compartment,
\begin{equation}
	\kappa=- \frac{1}{V_i}   \frac{\partial V_i }{\partial p_i}   
\end{equation}
is the compressibility, inverse of the bulk modulus $K$,  
and $V_{ei}$ is the volume of the compartment in the absence of osmosis. The corresponding number of molecules  in this volume is $N_{ei}$.

The transition of  molecules  is accompanied with the change of the volume of the compartments. 
The mean  distance between  molecules $r_i$, the number of transferred molecules $\Delta N_i$ and the change of the volume  $ \Delta V_i$ are related by equation,  
\begin{equation}
\label{VN}
V_i=V_{ei} + \Delta V_i = (N_{ei} + \Delta N_i) r_i^3.
\end{equation} 
The compressibility is the relative change of the liquid volume with a fixed number of molecules due to the change of  pressure.
In our case  the starting volume of $N_i$ molecules 
\begin{equation}
\label{VeNi}
  V_{eN_i}=(N_{ei} + \Delta N_i)r_e^3
\end{equation}
   has been changed to $V_i$, due to the change of the pressure $p_i-p$.
The compressibility now reads,
\begin{equation}
	\kappa=- \frac{1}{V_{ei}}   \frac{ V_i-V_{eN_i} }{ p_i-p}.
\end{equation}
By means of (\ref{VN}) and (\ref{VeNi}) the compressibility  becomes
\begin{equation}
	\kappa= \left( 1+\frac{\Delta  V_i}{V_{ei}}\right)   \frac{ \left(\frac{r_e}{r_i}\right)^3-1   }{ p_i-p}.   
\end{equation}
Keeping only the linear term in $r_e-r_i$ we get 
\begin{equation}
\label{kappa}
\kappa =3 \frac{r_e-r_i}{r_i(p_i - p)} \qquad \qquad \begin{array}{l c r} p_i > p & & r_i < r_e \\ & & \\ p_i < p & & r_i > r_e  \end{array}.
\end{equation}

On the other hand the change of the  pressure  is a function of the relative change of the volume. In the linear approximation, taking into account that the  positive change of the volume leads to the increase of the pressure,  we write 
\begin{equation}
\label{pi}
p_i - p_a = c \frac{\Delta V_i}{V_{i}}.
\end{equation}
The constant $c$ is determined by elastic properties of membrane and will be calculated in the next section.
In the following text slight difference between $V_{i}$ and $V_{ei}$ plays no role.
For the sake of simplicity hereinafter we write $V_{i}$ instead of $V_{ei}$

The pressure difference between the compartments is 
\begin{equation}
\label{Deltap}
\Delta p=p_2 - p_1 = c \Delta V \left(\frac{1}{V_{1}}+ \frac{1}{V_{2}} \right).
\end{equation}
 
 By means of (\ref{deltaF1}), (\ref{kappa} and (\ref{pi}) the corresponding free energy change is,  
\begin{equation}
\label{Fw}
\Delta {\cal F}_w  = \frac{\kappa c^2}{2} \Delta V^2 \left (\frac{1}{V_{1}} + \frac{1}{V_{2}}\right ).
\end{equation}
As it is expected the stable state, the state of minimum free energy,  of the water in  both compartments  is the state of the zero change of volume.
Then according to Eq. (\ref{pi}) pressures in both compartments are equal.

\subsection{Elastic energy stored in semipermeable membrane}
\label{b}

\begin{figure}[ht]
\centering
\resizebox{8cm}{4cm}
{\includegraphics{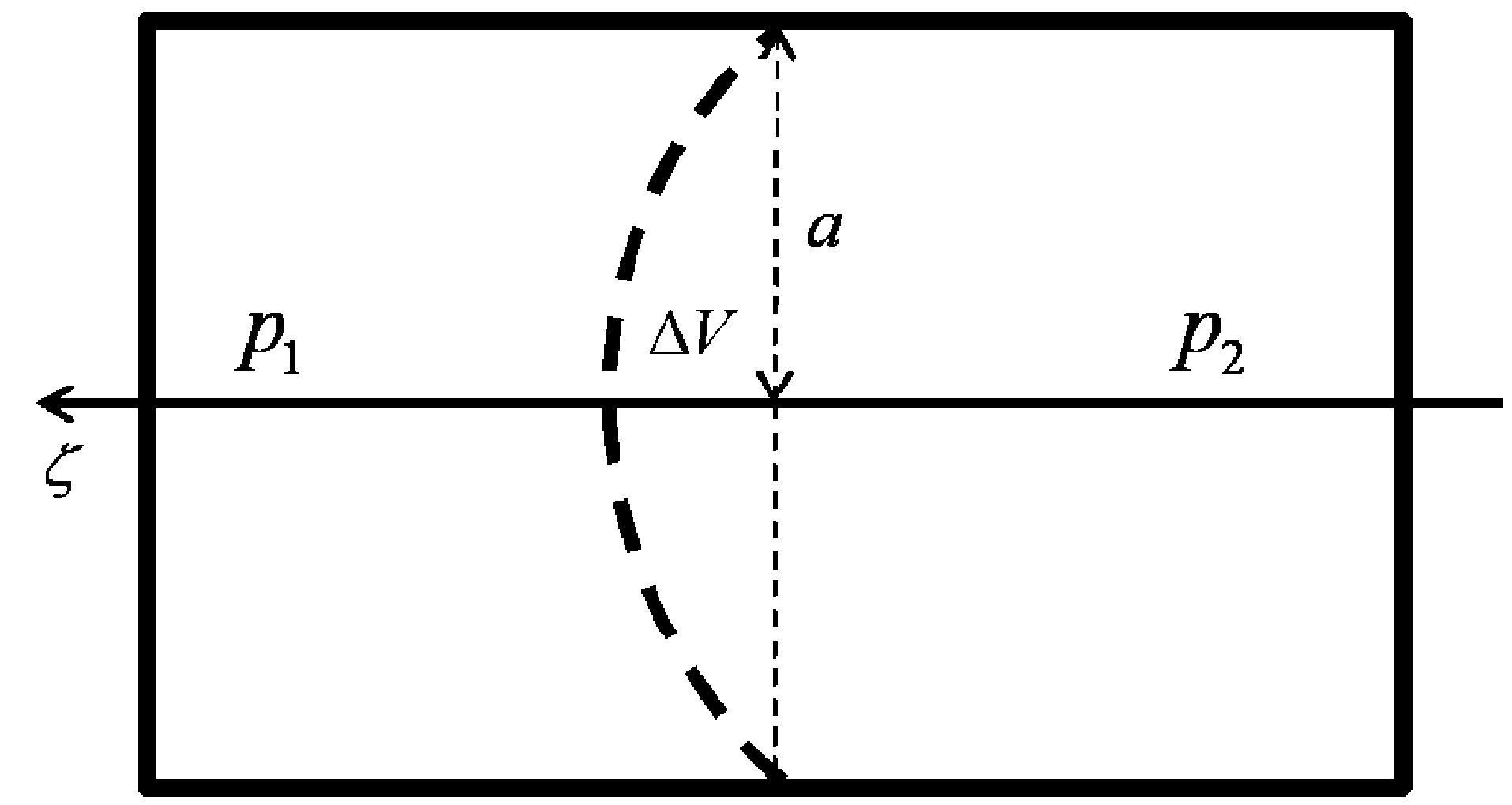}}
	\caption{\label{Figure3} The deflection of the elastic membrane  under  the pressures difference. }
\end{figure}

The pressure difference between compartments stresses the elastic  semipermeable membrane (see Fig. \ref{Figure3}). 
  Elastic energy stored in the membrane should  be   added to the water  free energy change (\ref{Fw}) in order to complete the 
 energy contributions to  free energy change.

A problem of a small deflection of the membrane
due to pressure difference is similar to  the problem  of a deflection of the membrane due to its own weight. Latter is   the standard problem of the theory of elasticity \cite{landauTE}. In the case of a circular plate 
(see Fig.\ref{Figure3}) the deflection  is given by expression,
\begin{equation}
	\label{def}
	\zeta=3 \Delta p \frac{(1-\sigma^2) }{16h^3E}\left(a^2-r^2\right)^2.
\end{equation}
 Here $E$, $\sigma$, $r$ and $h$ are the Young's modulus, Poisson's ratio,  a distance between  the centre of the plate  and some point of the plate and a thickness of the plate, respectively,

 The change of the volume of  the compartment is
 proportional to   the pressure difference. An elementary calculation gives  
\begin{equation}
	\label{deltap}
	\Delta p = \frac{16Eh^3}{ \pi a^6(1-\sigma^2)} \Delta V.
\end{equation}
Eqs. (\ref{Deltap}) and (\ref{deltap}) define the constant $c$, 
\begin{equation}
	\label{c1}
	c  =\frac{16Eh^3}{\pi a^6(1-\sigma^2)}\frac{V_1V_2}{V} .
\end{equation}

Elastic energy stored in plate \cite{landauTE} is given by expression,
\begin{equation}
\label{energy}
E_{el}=\frac{Eh^3}{24(1-\sigma^2)} \int \int \left\{ \left(  \frac{\partial^2 \zeta}{\partial x^2}+ \frac{\partial^2 \zeta}{\partial y^2}\right)^2+2(1-\sigma)\left[ \left(  \frac{\partial^2 \zeta}{\partial x \partial y}\right)^2- \frac{\partial^2 \zeta}{\partial x^2} \frac{\partial^2 \zeta}{\partial y^2} \right] \right\}dxdy.
\end{equation}
In the case of the axial symmetric plate the
foregoing expression  written in  the cylindrical coordinates becomes 
\begin{equation}
\label{elastic}
E_{el} = \frac{\pi}{12}\frac{h^2 E}{1-\sigma^2}\int_0^R \frac{\partial^2 \zeta}{\partial r^2} \left( \frac{\partial^2 \zeta}{\partial r^2}+\frac{2 \sigma}{r} \frac{\partial \zeta}{\partial r}\right) r dr  .
\end{equation}
After a  long but otherwise   straightforward calculation we get from Eqs. (\ref{def}) and (\ref{elastic}),
\begin{equation}
	\label{elastic1}
	E_{el} =\frac{\pi R^6  \left( 1-\sigma^2 \right)} {32 h^3E}  (\Delta p)^2.
\end{equation}

Elastic energy of the  membrane in terms of the change of volume is given by expression
\begin{equation}
	\label{elastic3}
	E_{el}=  \frac{8h^3E}{\pi( 1-\sigma^2) a^6}(\Delta V)^2.
\end{equation}

\subsection{Entropy }
\label{c}

An expansion  of the solute  into the compartment (see Figs. \ref{Figure2}a and \ref{Figure2}b) modifies both  energy and entropy of the system.
In order to make the calculation as simple as possible, but without loss of any relevant effect, we assume 
 that interaction between solute and water molecules is similar to the interaction between  water molecules themselves. This assumption allows us to  discard the change of internal energy due to the mixing water and solute. In other words 
solute-water, water-water interactions are taken to be   equivalent. 
However we cannot ignore the change of entropy due to the space delocalization of the solute. We use the free gas expression for  entropy   to describe the increase of   entropy due to the spread of solute into the compartment  
( Figs. \ref{Figure2}a and \ref{Figure2}b),
\begin{equation}
\label{entr}
 S_s  =n_s V_2R \ln (V_2 + \Delta V) \approx n_sV_2R \ln V_2 + n_sR \Delta V.
\end{equation}
Here $n_s$ is the molar concentration of solute and $R$ is the gas constant.

\subsection{Total free energy }
\label{d}

The expansion of the solute  in the compartment is accompanied with the change of free energy of water (\ref{Fw}) elastic energy of the membrane (\ref{elastic3}) and increase of  configuration entropy (\ref{entr}).
Total free energy ${\cal F} = {\cal F}_{w0} + \Delta {\cal F}_w +E_{el} - TS_s$, where ${\cal F}_{w0}$ is  equilibrium free energy of  water without any solute, reads
\begin{equation}
\label{free11}
{\cal F} = {\cal F}_{w0} +  \frac{Eh^3}{\pi a^6(1-\sigma^2)} \left[\kappa\frac{128Eh^3}{\pi (1-\sigma^2)a^6} \frac{V_1V_2}{V} +8 \right](\Delta V)^2 - n_sV_2RT \ln V_2- n_sRT \Delta V. 
\end{equation}

\subsection{Osmotic pressure}
\label{osmotic}

The minimum of the free energy  ($\partial {\cal F}/\partial \Delta V = 0$) determines equilibrium change of the volume,

\begin{equation}
\label{DeltaV}
\Delta V = n_sRT \frac{\pi a^6(1-\sigma^2)}{2Eh^3\left[\frac{128 \kappa Eh^3}{\pi a^6(1-\sigma^2)} \frac{V_1V_2}{V} +8 \right] }.
\end{equation}

Equation (\ref{pi}) is the general expression that relates  change of the volumes of the compartment and   changes of  pressure. 
In the case of the osmosis  the  solute creates change of pressure in the compartment ( see Eqs. \ref{pi} and \ref{DeltaV}).
 We call this change the partial osmotic pressure, $\Pi_i$. Then Eq. (\ref{pi}) reads, 
  \begin{equation}
\label{Pii}
p_i=p_a+\Pi_i.
\end{equation}
The osmotic pressure is defined as the 
difference between pressures in compartments
 \begin{equation}
\label{Pi}
\Pi=p_2-p_1=\Pi_2-\Pi_1.
\end{equation}
By means of  Eqs. (\ref{pi}), (\ref{c1}) and  (\ref{DeltaV}) the  osmotic pressure becomes,
\begin{equation}
\label{osmoticp}
	\Pi=p_2-p_1=\Pi_2-\Pi_1 =\frac{8n_sRT}{\frac{128 \kappa Eh^3}{\pi a^6(1-\sigma^2)} \frac{V_1V_2}{V} +8  }.
\end{equation}
The osmotic pressure $\Pi$ comes due to the decrease of the pressure in the water compartment and the increase of the pressure in the compartment with the solution. 
The first and  the second  term in the denominator of the osmotic pressure (\ref{osmoticp})  are proportional to  water and membrane free energy contributions (see Eq. \ref{free11}).  
The value of the ratio $\eta$  between  these two terms 
\begin{equation}
\label{composition}
\eta=\frac{\frac{16  Eh^3}{\pi a^6(1-\sigma^2)}}{  \frac{KV}{V_1V_2} }  , 
\end{equation}
determines the increase of  internal energy  of water 
 relative to  elastic energy  stored in  the membrane. If this ratio is much less than one $\eta \ll 1$,  due to  the high bulk modulus of water $K$ relative to the modulus of elasticity  $E$  and/or due to
 geometrical parameters    of the  membrane $h^3/a^6$, and volumes of compartments $V_1V_2/V$, 
osmotic pressure is given just by van't Hoff equation,
\begin{equation}
 \label{OP}
	\Pi=n_sRT.
\end{equation}
In the opposite case $\eta \gg 1$ the increase of internal   energy  is mainly stored in water and the osmotic pressure reads, 
\begin{equation}
\label{osmoticp2}
	\Pi=n_sRT\frac{\pi a^6(1-\sigma^2)}{16 \kappa Eh^3} \frac{V}{V_1V_2}.
\end{equation}
 In short the small product  of the membrane modulus of  elasticity  and the geometrical parameters of the membrane and compartments in comparison with the bulk modulus of water     implies that  the change 
 of internal energy  is mainly elastic energy of  the membrane. In the opposite case  the increase of internal  energy   is mainly stored in the compressed solution and tensed water. This result is equivalent to the fact that 
  product of energy   stored in a spring and a corresponding force constant  is constant  for springs connected in series and loaded with  opposite forces.

 We now turn to the partial osmotic pressures. They can be expressed in terms of osmotic pressure and relative volumes of  the compartments, 
\begin{eqnarray}
\label{p1}
\Pi_1 =- \Pi\frac{V_2}{V}, \\
\label{p2}
\Pi_2 = \Pi\frac{V_1}{V} . 
\end{eqnarray} 
The distribution of the osmotic pressure between compartments depends on their relative  volumes. In the  case of the  small
 water compartment, $V_2 \approx V $, there is no change of the pressure in the compartment with the  solution while 
 pressure in the water compartment is decreased for the total osmotic pressure $\Pi$. In the opposite case,  $V_1\approx V$,  
 there is no change of the pressure in the water  compartment  while the
 pressure in the compartment with the solution is increased for the value of  osmotic pressure $\Pi$.

In the next sections we argue that the osmosis under  constant volume plays a major role in the problem of water uptake in tall trees. 
The xylem and phloem with ray cells are counterparts to the box compartments with water and solution, respectively. 
In order to estimate contributions of terms in equation  (\ref{composition}) we put $V_2 \approx V$. 
The upper bound value of the  Poisson's ratio is $0.5$. In the most cases  this ratio is close to $0.3$. Hardwoods \cite{poisson} are not exception. 
The bulk modulus of water ($\approx 2 \cdot 10^{9} Pa$) and  typical values of radial and tangential  modules of elasticity of hardwoods   ($\approx   10^{9} Pa$) are known parameters \cite{poisson}. 
Assuming that a transverse dimension of the  tracheary element is  one order of magnitude higher that  thickness  of  walls between cells of xylem  parenchyma and tracheary elements ($h/R \approx 0,1$)
and one order  of magnitude less than a longitudinal dimension of  the tracheary element ($L/R \approx 10$)  we get
\begin{equation}
\label{condition}
\eta \approx \frac{16  E}{K \pi(1-\sigma^2)} \left(\frac{h}{R}\right)^3 \frac{L}{R} \approx 0.025.
\end{equation}
The small value of parameter $\eta$ shows that the osmotic pressure between the xylem and phloem is given by van't Hoff's equation. Due to the much larger volume of the phloem with ray cells 
 in comparison with the  xylem it comes out form Eqs. (\ref{p1}) and (\ref{p2}) that the pressure in the xylem is decreased for almost the total value of osmotic pressure. On the other hand the  pressure increase in the phloem, due to osmosis, is negligible. 
This result  stays in  contrast to the result of  the osmosis under  constant pressure. Latter is  characterized by an increase of the  pressure in the solution 
 just for the  osmotic pressure given by van't Hoff equation. However  in the case of the osmosis under  constant volume there is no  increase of the pressure 
  in the  solution  if the  volume of compartment with the  solution is much larger than one with water.

  \section{Osmosis  in  closed  tall vertical tube}
  \label{tube}
   
\begin{figure}[ht]
\centering
\resizebox{10cm}{11cm}
{\includegraphics{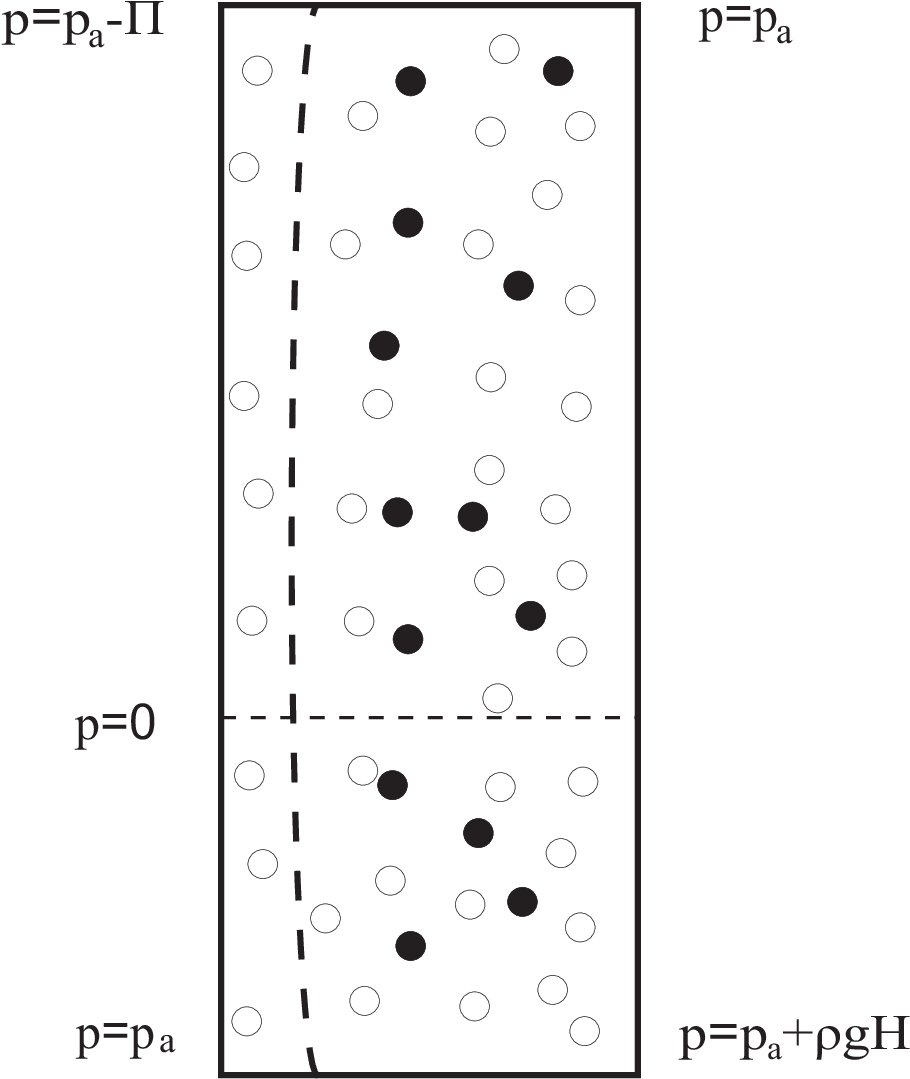}}
\caption{\label{Figure4} Water and the solution separated in two compartments by the semipermeable membrane in the tall tube. Values of pressures in this picture correspond to the 
	 case of small water compartment, in comparison with compartment with solution, and   hydrostatic condition. } 
\end{figure}

 We consider another example of the osmosis under   constant volume. 
    The closed tall tube is divided into two compartments by the  vertical semipermeable membrane as it is depicted in Fig. \ref{Figure4}.  The volume of the water compartment is much smaller than one with the solution. As we have seen in the previous section, there is very small increase  of pressure in solution compartment due to the osmosis. The whole effect of the osmosis 
    is present in  the water compartment.

    The pressure in the water compartments reads, 
    \begin{equation}
    p_1=p_a- \Pi(z)+\rho g(H-z).
    \end{equation}
   Here  $\Pi(z)$ is the osmotic pressure  and  $H$ is the height of  tube. 
      Depending on the solute concentration and a value of vertical coordinate $z$ water can be either under the positive or negative pressure.
           Here,  we are interested in moderate concentrations that 
  do not 
      depend heavily  on $z$ coordinate. Then the  lower part of tube is under the positive and the upper part of tube under the negative pressure (see Fig.\ref{Figure4}), respectively. 
   The hydrostatic situation,  with constant concentration of solution   is depicted in Fig.\ref{Figure4}.
   An assumption that  pressure at the bottom in the water  compartment water is equal to atmospheric pressure implies that the osmotic pressure  compensates the  hydrostatic pressure of the water 
    column.

 We ask how large hydrostatic pressures can be compensated by osmotic pressures that are  generated by measured values of photosynthetic products in the phloem.
  A sugar concentrations in the phloem from  $0.2$ to  $0.7$ mol/l are reported in the reference   \cite{Nobel}. According to Eq. (\ref{OP}) these concentrations of the solute can compensate  hydrostatic pressures of    water columns with a height in range between  $48m$ and $168m$. 
   However  in the hydrodynamic case one needs to add the viscosity part of the  pressure  gradient  to the hydrostatic part of  the pressure gradient.
   It is argued  in  the reference \cite{Lampinen} that the viscosity part of the pressure gradient for Scots pine of  $-1.4 \cdot 10^4 Pa/m$ should be compensatory in order to keep stationary 
    sap flow in the xylem with velocity of $5cm/h$. 
   According to these data the total gradient of pressure is $-2.4 \cdot 10^4 Pa/m$.  Under these   conditions, which are  typical for Scots pine,   reported range of the sugar concentration can  sustain stationary  upward water flow through the xylem in trees within  the range of  a height  
    between $20m$ and $70m$.

 \section{Modifications of containers that do not affect osmosis under constant  volume}
 \label{4}
 
\begin{figure}[ht]
\centering
\resizebox{15cm}{11cm}
{\includegraphics{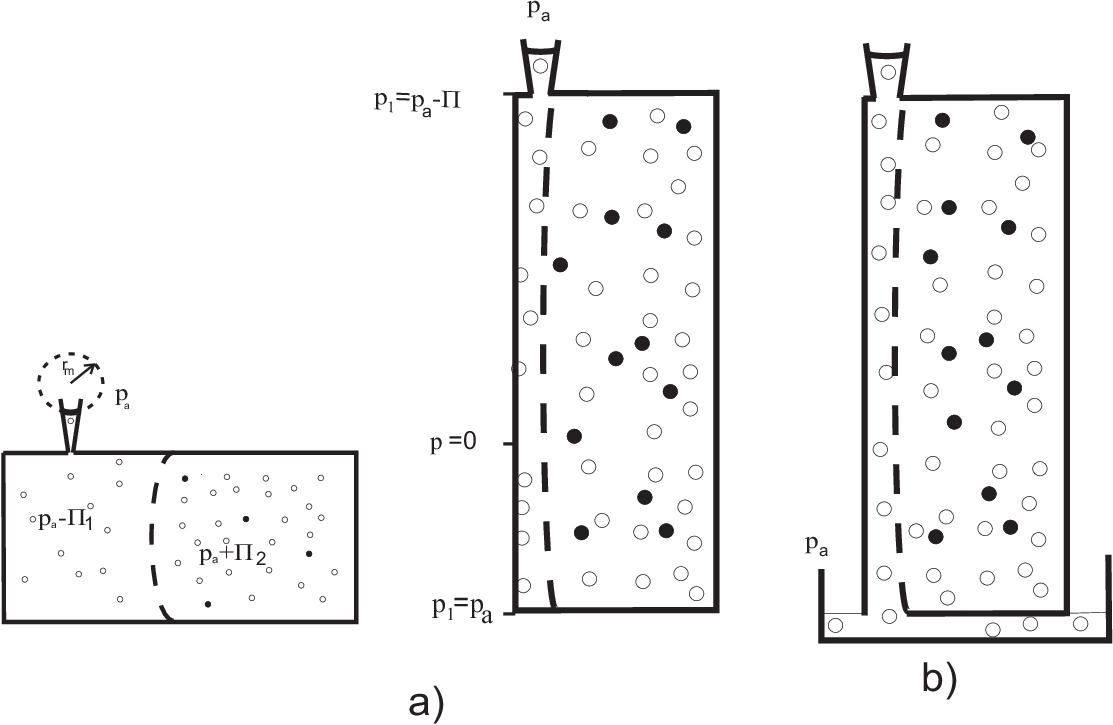}}
	\caption{\label{Figure5} a)The microscopic funnel attached on the water compartment has no effect on the osmosis under constant volume. b) The tube  is immersed into the  bucket with water and the lower base of the water compartment is removed.} 
\end{figure}

  Some modifications can be done on  containers    that do not  spoil the  osmosis under  constant volume.
  Let us assume that a microscopic small funnel is attached to the water compartment (Fig. \ref{Figure5} a).
   The meniscus compensates  the pressure difference between the atmospheric  pressure  and  the pressure  within the  water compartment. Then the partial 
    osmotic pressure is equal to the 
     surface tension contribution to the pressure,
       \begin{equation}
   \Pi_1 =\frac{2 \alpha \cos(\Theta)}{r_m}.
    \end{equation}

 As in the case of the  box  the microscopic funnel on the part of the upper  base of the closed vertical  tube over the water compartment (Fig. \ref{Figure5} a) does not spoil the osmosis under constant volume. 
Since we have assumed the atmospheric pressure at the bottom of the water compartment the immersion of the  tube into the bucket with water 
 and removing  the lower  base of the compartment with   water   does not pertubate thermodynamic states of liquids in tube. In other word the state of 
   liquid is the osmosis under constant volume (see Fig. \ref{Figure5} b ) .

\section{Stability of  osmosis at constant volume. Cavitations and refilling}
\label{5}
  
   \begin{figure}[ht]
\centering
\resizebox{8cm}{14cm}
{\includegraphics{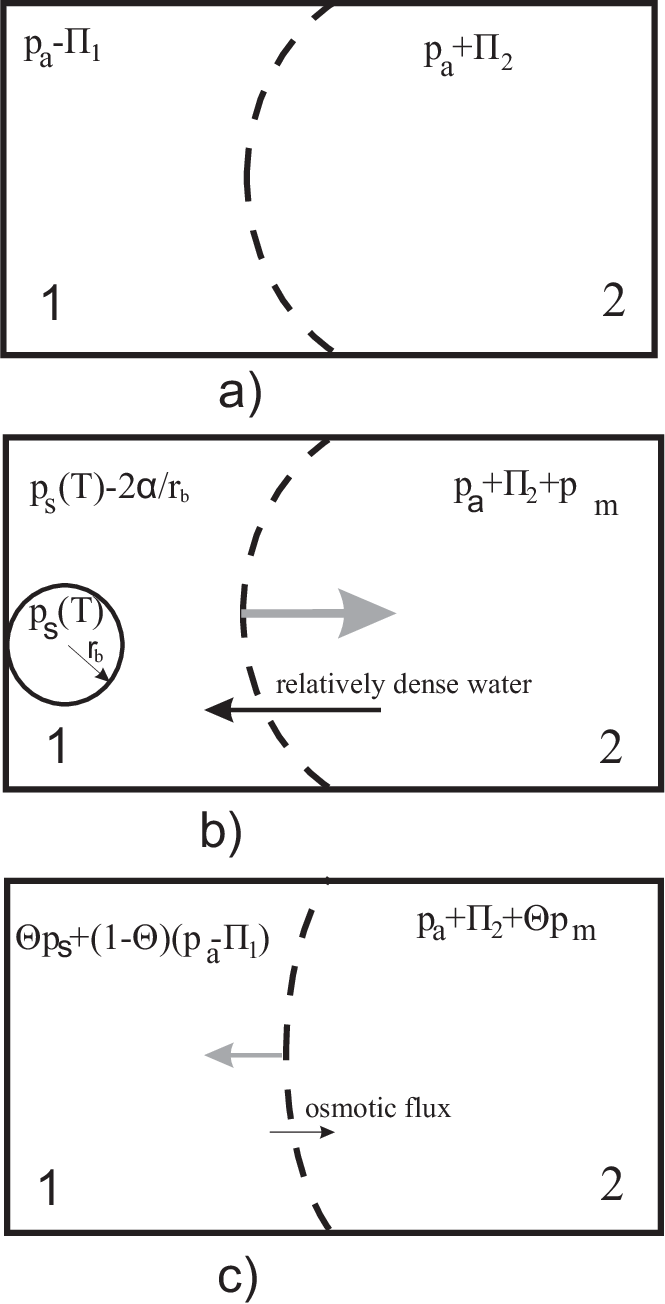}}
	\caption{\label{Figure6} 
	 a)Water is under the negative pressure (metastabile state). b)Cavitations  occurs and
adhesion forces stick bubbles to the inner walls of water compartment.   The membrane relaxes increasing  the  pressure in solution  by $p_m$. Due to the increase of the pressure in the compartment with the solution   relatively  high dense water flows  from the
solution to water.
 c) Cavitation is repaired and osmosis takes place. Pressure in water and solution approach to the values in metastabile state  ($\Theta \rightarrow 0$). Gray arrows indicate  directions of a membrane deformation.}
\end{figure}

 Water in the osmosis under constant volume is under the negative pressure and cavitations take place. 
 We predict here that the elastic membrane  can, under certain circumstances, refill the cavitated compartment.

 Suppose that the osmotic  state with the pressures given by Eqs. (\ref{p1}) and (\ref{p2}) is established. The elastic membrane is under the tension in one part due to the high pressure in the compartment with solution and in other part
  due to the  adhesive forces between the tensile  water and the elastic membrane (Fig. \ref{Figure6} a). 
 Due to cavitations   bubbles of the saturated  vapor appear (Fig.\ref{Figure6} b)  and    the    pressure in the water compartment  becomes equal to the pressure of saturated vapor  reduced by the surface tension,  $p_s(T)-2\alpha/r_m$,  of a bubble.  
    The tension due to  
 adhesive forces between  water and the  membrane vanish. The membrane relaxes pushing  relatively high density water from the compartment with solution to the  compartment with  water  (Fig.\ref{Figure6} c). 
  If the   compartment with water   is shallow and if/or the membrane area is  large  enough, the increase of the pressure in the water compartment  can be 
  high  enough  to cause the collapse of  bubbles into the liquid. 
  Once the liquid phase is established the osmosis takes place creating again the negative pressure in  the water compartment (Fig. \ref{Figure6} c).
  The  system acts like a bistable devices. It oscillates between  the osmotic  and  cavitated  state. 
  Transitions between these states  occur on different time scales. The osmosis is  the diffusion and it is very  slow process in comparison with the relaxation of the membrane.
  The frequency of oscillations is dominantly determined with the osmotic process.

\section{Water uptake in tall trees and cavitations repair}
\label{xylem}

The long vertical tube with the funnel at the top immersed into a bucket with water is the crude  model of the vascular system in tall trees
 (see Fig.\ref{Figure5} b). The compartments with water and solution  are analogous to  the xylem
 and phloem with ray cells, respectively. In  contrast to the CT theory the   negative pressure within the xylem is  mainly 
 the result of the osmosis under  constant volume. 
  As we have shown in Section \ref{4}  the existence of funnel like  pores in walls of  mesophyll cells  is consistent with the osmosis under constant volume. In our model evaporation is  a  drain of water.
  It  is not only drain of water in trees. A minor drain is due to photosynthesis.  Both of these  processes decrease the
   pressure (making it  more negative) in the foliage part of the xylem.
   To be more specific a radius of water meniscus in a pore of a mesophyll cells decreases due to evaporation reducing in this way a pressure below a meniscus. 
  In the case of the  photosynthesis the rise of the concentration of photosynthetic products   in the phloem increases the osmotic pressure. As we have seen in Section \ref{osmotic}
   the increase of the osmotic pressure reflects solely  in  the decrease of the pressure in the xylem. Neither the evaporation nor photosynthesis spoils the osmosis 
    under constant volume.  The only effect of these two processes is an enhancement of pressure gradient in the xylem  which uptakes water.

   Osmotic pressures are usually much higher than the atmospheric pressure.
   Having in mind that the phloem with ray cells  is much larger than the xylem
   the negative pressure  in xylem close to foliage, in hydrostatic condition,  is equal to osmotic pressure given by van't Hoff equation (see Eqs.\ref{p1},\ref{p2},\ref{OP} and Fig. \ref{Figure5}).  
   In    reference \cite{Scholander}  
    close agreement has been found
between the negative hydrostatic pressure in the xylem measured by the pressure  bomb \cite{PressureBomb} and the
osmotic pressure obtained by inserting into van't Hoff equation (\ref{OP}) concentrations of photosynthetic products and their derivatives in ray cells.
Concentrations of photosynthetic products and their derivatives were determined by measuring the change of the water freezing point.
This  statement is just the content of Section \ref{tube}.
In other words    our proposal that the osmosis under constant volume between xylem and phloem  with ray cells 
is responsible for the water uptake in the  xylem is in accordance with the Scholander's  experimental result \cite{Scholander}.

The main objection to the  CT theory comes from  the fact that it cannot account for the stability of the tensile sap in the xylem \cite{Zimmerman}. 
 There must be some mechanism that stimulates  a process reverse to  cavitations. 
In the previous sections we have seen that the osmosis under constant volume is a  good candidate for  cavitations repair if the water compartment is not too broad and if the 
membrane is not too  small.
The long vertical tube is not a suitable geometry for  cavitations repair.  
 Biological evolution has resolved uninterrupted water uptake 
 compartmentalization of the  xylem into the system of interconnected tracheary
elements.

 Pores in the pit membrane are few nanometers wide and the membrane is impermeable for  products of photosynthesis and their derivatives. The presence of   products of photosynthesis and their derivatives 
in the xylem parenchyma induces the osmosis  between  the xylem and xylem parenchyma.

\begin{figure}[ht]
\centering
\resizebox{16cm}{13cm}
{\includegraphics{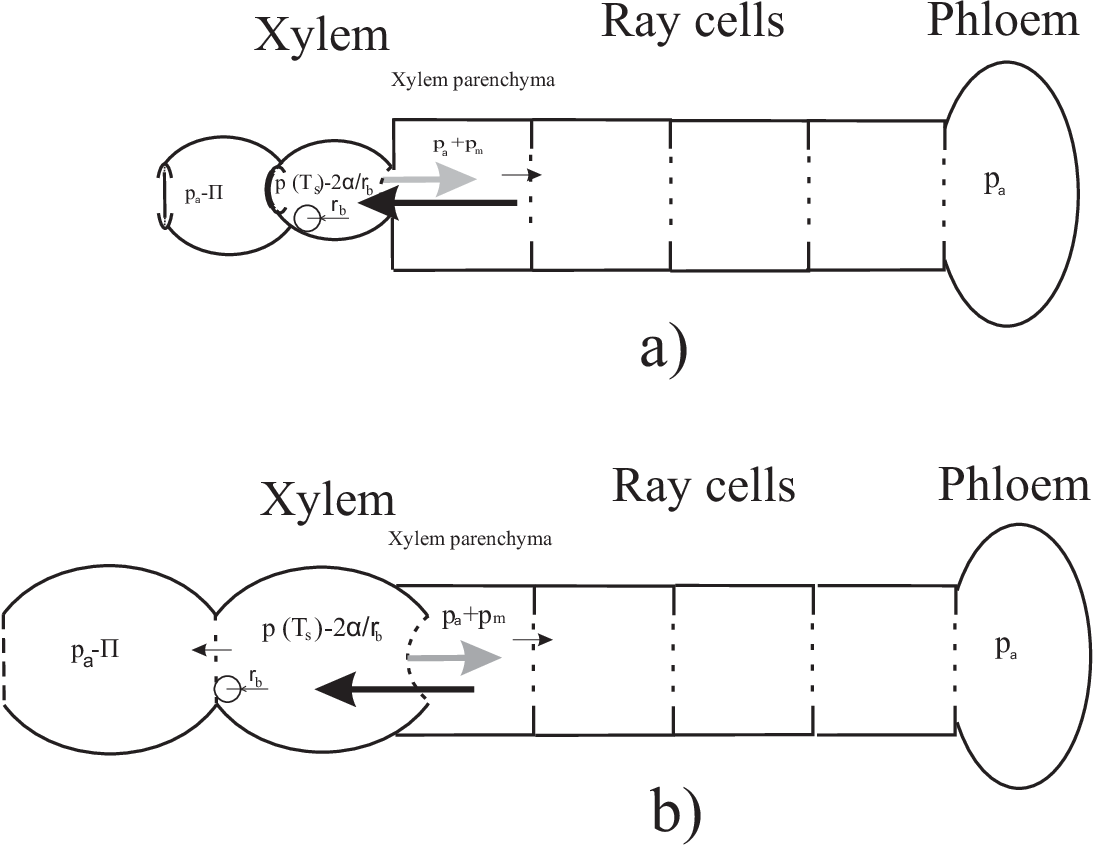}}
	\caption{\label{Figure7}   Relative radial  position of the
	 xylem, ray cells and phloem. Bubbles of the saturated  vapor are in the cavitated tracheary elements.  $p_m$ is the increase of  pressure in xylem parenchyma due to the relaxation of elastic pit membrane (see Fig.\ref{Figure6}). Large and small black arrows directed from the xylem parenchyma to the xylem and phloem 
	 denote corresponding large and tiny water fluxes.  The gray arrow shows the direction of the pit membrane relaxation. a) Due to the bordered pits between tracheids there is no water leakage from the cavitated tracheid into neighboring ones.   b) Vessels are radial interconnected by pit membranes. Menisci in pores of the  pit membrane   prevent cavitations spread into neighboring vessels but do not prevent the leakage of water.   An arrow between vessels indicated the water leakage into vessels with the tensile sap.   Values of  pressures pertain  to the top of tree (see Fig. \ref{Figure4}). }
\end{figure}

 Walls between a tracheary element and  cells of the xylem parenchyma are  tensed mainly  due to the  negative  pressure in tracheary elements.
 Once  cavitations take place the negative  pressure disappears and  walls  relax  increasing the pressure in the xylem parenchyma (see Fig.\ref{Figure7}). 
   In this way 
  cells of xylem parenchyma  become   sources that distribute water in a radial direction.   
   Two  radial water fluxes (see Fig. \ref{Figure7}) in the opposite directions, to the xylem and phloem,  are established.  The flux is proportional to the pressure gradient and the hydraulic conductivity. 
 From the geometrical point of view both of these quantities are inversely proportional to the 
  distance between the source and sink. 
 Due to the  much larger distance between  the xylem parenchyma and  phloem in comparison with the distance between the xylem parenchyma and xylem   (see Fig.\ref{Figure7}a) the  geometrical contribution to the radial   water flux  to the phloem is much smaller than  one to the xylem. The hydraulic conductivity of the water pathway to the tracheary  element  is additionally enhanced by abundant quantity of aquaporin proteins
 (proteins acting as water channels) 
  in the xylem parenchyma  \cite{2amerikanca2,Barrieu}. 
  
Tracheids and vessels  have different geometry and anatomy structures. Tracheids are  narrower than vessels \cite{TyreeBook}  and they have bordered pits in side walls in contrast to  pits in vessel side 
walls 
\cite{PlantPhysiology}  (compare Figs. \ref{Figure7}a and \ref{Figure7}b) . In the case of water flow from the xylem parenchyma into the  tracheid, due to the pit membranes relaxation, there is no leakage of  water from the tracheid into  the rest of   xylem. This is not a case with vessels. Due to  pits in side walls  cavitated  vessel   is the water  sink for the xylem parenchyma and water  source for the rest of low pressure 
 xylem. Cavitations will be repaired if the incoming water  flow  overwhelms the outgoing water flow  (see Fig.\ref{Figure7}b). 
 Both of these structural differences  make the tracheid superior structure  in comparison with  the vessel regarding  cavitations repair. 
 Superiority of the tracheid in  refilling explains why six \cite{wiki} of ten tallest trees in word are conifers, although the number of different angiosperms  \cite{PlantPhysiology} ($\approx 250000$) overwhelms
  the number of conifers (less than $ 700$).

  In short, the  
  extended vascular transport system in trees can be modeled, at the first step,  with the long tube of the constant volume  divided into two 
   main compartment by the semipermeable membrane. Draining of  water by the evaporation and photosynthesis increases  the pressure gradient in the xylem which moves water  upward.
   The model of the long tube does provide enough negative pressure that can lift water from the root to the top of tree, but 
   due to the metastability of  water under the tension  the   water uptake   would be interrupted by cavitations, at least temporary.  Refilling of the broad  cavitated  compartment   is highly improbable.
   Uninterrupted water uptake  the biological   
      evolution has resolved by  compartmentalization of the   xylem into the system of interconnected  tracheary  elements.

\section{ Cavitations repair and  the second law}
\label{SecondLaw}

In the Section \ref{5}  we propose the mechanism that could account for cavitations repair.
From the thermodynamic point of view the box is the system in  contact with a thermal bath of the constant temperature. 
The elastic membrane does work. At the first sight one can object that work is done solely by absorption of  heat from one heat bath. 
Than the suggested mechanism would  in variance  with the second law of thermodynamics. However this is only an
apparent contradiction.

The proposed mechanism of cavitations repair  takes for a granted    that the reflectivity of the semipermeable membrane is $100 \% $. 
In reality this is not a case. 
 Due to the nonideal reflectivity of the semipermeable membrane the solute steadily diffuses through the  membrane converting water  into the solution. The difference of solute concentrations in compartments, the thermodynamic force of the osmosis, 
         steadily decreases. 
         A small osmotic thermodynamic force cannot establish the state of negative pressure. In this way    water uptake in xylem  is disabled. Vessels that do not take a part in the water transport form the heartwood, the
          inner hydraulically nonconducting    part of the stem.  
         The problem of relatively  short lifetime of functioning  vessels the biological evolution has solved creating each year a new layer of secondary xylem, ring. These functioning vessels form 
        the  sapwood, 
          outer hydraulically conducting part of the stem.

\section{Concluding remarks}
\label{8}
The standard textbooks on  thermodynamics  \cite{zemansky,kubo,guggenheim}  introduce the pressure within kinetic theory of gases to describe the free gas state.
 Due to the weakness of cohesive forces in  gases these systems cannot be under the tension. 
  This property of the gases  has served as the  basis for an extrapolation of the  pressure as the positive definite value to other systems. 
  However the extension of  the concept of pressure to  liquids under tension  leads to the notion of the   negative pressure as the well defined  physical quantity.

  We argue  in this paper that water in osmotic process under   constant volume can be in the state of the negative pressure. Generally the state of negative pressure in liquids  is  metastabile. Cavitations occur and the phase transition from the tensile state to the  state of coexisting liquid and bubbles of saturated vapor occurs.
    Our qualitative analysis of the osmosis under constant volume  in Section \ref{xylem} predicts that energy stored in the elastic membrane can induce   refilling,  
    the   process reverses  to  cavitations.  Once  cavitations are repaired the osmosis takes place driving the water into the tensile state again.
     We have modeled the extended vascular transport system in trees  with the long tube of constant  volume  divided into two compartments by the semipermeable membrane. 
     The smaller compartment  with interconnected units    corresponds to  the xylem  composed of  
     tracheary elements. The main result of this paper is that  
     two  long standing problems in the tree physiology, water uptake and refilling  of  tracheary  elements, can be explained holistically within this model. We note that our explanation is in accord with M\"{u}ch  statement that no flow of solution in plants  is possible if it is not caused, or at least influenced, by osmosis \cite{munch}.

         In addition we have critically considered the  proposed mechanism of cavitations repair  from the thermodynamic point of view.
         The  blockade of the osmosis under constant volume   due to the  nonideal reflectivity of the membrane is related to the second law of thermodynamics. 
         The biological evolution has avoided the blockade of water uptake in the xylem due to the  diffusion of photosynthesis products and/or their derivatives into the xylem 
         by an annual growth of the functioning secondary xylem.

\begin{acknowledgments}

We acknowledge   dr.sc. Valerija  Dunki\'{c}, the member of Biological department of University of Split, who has helped us to become more familiar  with  vascular transport pathways in trees.

The present work was supported by the bilateral research project of Slovenia-Croatia Cooperation in Science and Technology, 2009-2010, Croatian Ministry of Science, Education and Sport, project no. 177-1770495-0476, and the financial support from the state budget by the Slovenian Research Agency (Program No. P1-0055).

\end{acknowledgments}

\end{document}